\documentclass[10pt]{article}

\usepackage[T1]{fontenc}
\usepackage[utf8]{inputenc}
\usepackage{listings}
\usepackage{amssymb}
\usepackage{amsmath}
\usepackage{amsthm}
\usepackage{float}

\usepackage{url}
\usepackage{hyperref}

\usepackage{xcolor}
\usepackage{verbatim}
\definecolor{keywordcolor}{rgb}{0.7, 0.1, 0.1}   
\definecolor{tacticcolor}{rgb}{0.0, 0.1, 0.6}    
\definecolor{commentcolor}{rgb}{0.4, 0.4, 0.4}   
\definecolor{symbolcolor}{rgb}{0.0, 0.1, 0.6}    
\definecolor{sortcolor}{rgb}{0.1, 0.5, 0.1}      
\definecolor{attributecolor}{rgb}{0.7, 0.1, 0.1} 

\newtheorem{theorem}{Theorem}

\title{Formalizing a Many-Sorted Hybrid Polyadic Modal Logic in Lean}

\author{
Alexandru Oltean
\and
    Bogdan Macovei
\and
   Ioana Leuștean \\
   {\small Faculty of Mathematics and Computer Science, University of Bucharest,}\\
{\small Academiei nr.14, sector 1, C.P. 010014,  Bucharest, Romania} \\
\small \texttt{\{andrei-alexandru.oltean, bogdan.macovei, ioana.leustean\}@unibuc.ro}}

\date{}

\begin{document}

\maketitle

\begin{abstract}
 We present a Lean formalization of a general hybrid modal logic with many-sorted signatures and polyadic modal operators. The system borrows ideas from both algebraic specification and dynamic logics, and is designed to serve as a uniform axiomatic foundation for specifying and verifying programming languages and security protocols. We expose a DSL for users to define languages and protocols as many-sorted signatures, specify the relevant domain-specific axioms, and reason about program executions or protocol runs. We provide a machine-checked proof of its soundness theorem and showcase the framework's versatility through several applications: an imperative programming language for code verification, a BAN logic and an epistemic logic used for security protocols, and the modal system S5. We have designed our formalization to be intrinsically sorted, that is, well-sorted formulas in the base language are well-typed terms in Lean. Thanks to intrinsic sorting, all domain specific applications can be easily embedded in our framework via the DSL, at no additional syntactic overhead required for the user to prove.
\end{abstract}

\section{Introduction}\label{introduction}


We present a Lean \cite{lean4} formalization of a general hybrid modal logic with many-sorted signatures and polyadic modal operators. The system borrows ideas from both algebraic specification and dynamic logics, and is designed to serve as a uniform axiomatic foundation for specifying and verifying programming languages and security protocols. Our approach embeds programming languages, their operational semantics, and correctness properties into a single logical framework, which is proved to be mathematically sound in the proof assistant's dependent type theory. Lean's modularity and extensibility make it a natural fit for this architecture: we expose a DSL for users to define languages or protocols as many-sorted signatures, specify the relevant domain-specific axioms, and reason about program executions or protocol runs entirely within the same base logic.

Our formalization focuses on the many-sorted hybrid modal logic $\mathcal{H}_{\bf \Sigma}(@, \forall)$ defined in \cite{tableaux}. We implement the syntax and semantics of this system, and we prove its soundness theorem for arbitrary axiomatic extensions. We implement three applications of our system, proving its ability to serve as a general verification framework. As a starting example, we embed an imperative programming language and its operational semantics into this logic: Plotkin's SMC machine \cite{plotkin}, following its hybrid modal treatment in \cite{tableaux}, which we formalize here. On this programming language, we use our modal proof system to derive Hoare-like inference rules and reason about program correctness. Second, we express different logics for the analysis about security protocols; drawing first from BAN logic \cite{ban}, and from earlier work on an epistemic logic with actions \cite{ramics}. Finally, we show an embedding of the basic modal language and its S5 proof system within our hybrid setting, demonstrating that \textit{fragments} of full hybrid expressivity can also be achieved through this formalization.
All of these applications are introduced by means of a custom DSL that allows the elaboration of Backus-Naur-like syntax into the data structures of many-sorted hybrid logic.


This paper is organized as follows. In Section \ref{prel} we shortly recall the logic $\mathcal{H}_{\bf \Sigma}(@, \forall)$. In Section \ref{syntax} we explain our design for mechanizing the syntax and proof system in Lean. In Section \ref{semantics} we show our implementation of the semantics and the soundness theorem. In Section  \ref{applications} we present the three applications we outlined above. The last section contains related results and conclusions.

All code presented in this paper is openly accessible in the following repository: \url{https://github.com/alexoltean61/msphml-lean}.

\section{Preliminaries: the System $\mathcal{H}_{\bf \Sigma}(@, \forall)$}\label{prel}

Hybrid modal logics originate in Arthur Prior's research on tense logics in the 1960s \cite{whatarehybrid}. Such systems are obtained from modal logics by adding a special kind of atomic symbols, called \textit{nominals}, as  syntactical references to the worlds of a Kripke model. Various systems of hybrid logics are known in the literature, examples being the treatment of the \textit{satisfaction operator $@_i$}, as well as \textit{state variables} and \textit{binders} ($\downarrow_x$, $\forall x$). We refer to \cite{whatarehybrid,handbook} for detailed accounts on this topic. In this section we recall the syntax and the semantics of the many sorted hybrid polyadic modal logic defined in \cite{tableaux}.

Let $S$ be a non-empty set of sorts. A \textit{signature with constant nominals} is a triple $(S, \Sigma, N)$, where we have $\Sigma = (\Sigma_{w,s})_{w \in S^+, s \in S}$, an $S^+\times S$-indexed family of countable sets of modal operators\footnote{Note that we disallow operators from having nullary arity. This choice was made for convenience, and all subsequent considerations can be adapted to fit the nullary case if required.}, and  $N = (N_s)_{s \in S}$, an $S$-indexed family of non-empty, countable sets of \textit{constant nominals}. 

Let ${\bf \Sigma} = (S, \Sigma, N)$ be a signature with constant nominals, and let $(PROP)_{s \in S}$, $(NOM)_{s \in S}$ and $(SVAR)_{s \in S}$ be $S$-sorted sets of \textit{propositional variables}, (non-constant) \textit{nominals}, and \textit{state variables}, respectively. For any $s\in S$, the well-formed formulas of sort $s$  are defined by:
   
        $\varphi_s := p \; | \; j \; | \; y \; | \; \neg \varphi_s \; | \; \varphi_s \lor \varphi_s \; | \; \sigma(\varphi_{s_1}, \dots, \varphi_{s_n}) \; | \; @_k^s \varphi_t \; | \; \forall x \varphi_s$,
   
    where $p \in PROP_s$, $j \in NOM_s \cup N_s$, $t\in S$, $k \in NOM_t \cup N_t$, $y \in SVAR_s$, $x \in SVAR_t$ and $\sigma \in \Sigma_{s_1...s_n,s}$. Derived propositional operators and the dual binder $\exists$ are defined as usual. For all operators $\sigma \in \Sigma_{s_1...s_n, s}$, the dual operator $\sigma^{\Box}$ is defined as $\sigma^{\Box}(\varphi_1,\dots,\varphi_n) = \neg \sigma(\neg \varphi_1,\dots,\neg \varphi_n)$. The language is \textit{hybrid}, due to the existence of nominals, constant nominals and state variables. We will refer to such symbols collectively as \textit{state symbols}.



The Kripke semantics for this  language is  based on frames and models, as usual but, in this setting, these are also many-sorted. 
    A \textit{$\bf \Sigma$-frame} is a tuple $\mathcal{F} = (W, (R_\sigma)_{\sigma \in \Sigma}, N^{\mathcal{F}})$, where:
    \begin{itemize}
        \item $W = (W_s)_{s \in S}$ is an $S$-sorted set of \textit{worlds}, each set assumed non-empty;
        \item $R_\sigma \subseteq W_s \times W_{s_1} \times \dots \times W_{s_n}$, for all $\sigma \in \Sigma_{s_1...s_n,s}$, is the \textit{accessibility relation};
        \item  $N^{\mathcal{F}} = (N^{\mathcal{F}}_{s})_{s \in S}$ with $N^{\mathcal{F}}_{s} : N_s \to W_s$ for all $s$. This is the \textit{valuation} of constant nominals. 
    \end{itemize}
\noindent If  $\mathcal{F} = (W, (R_\sigma)_{\sigma \in \Sigma}, N^{\mathcal{F}})$  is a $\bf\Sigma$-frame, then a $\bf\Sigma$-model based on  $\mathcal{F}$ is a tuple $\mathcal{M} = (\mathcal{F}, V)$, where $V =(V_s)_{s \in S}$ is an $S$-sorted \textit{valuation} function, with $V_s : PROP_s \cup NOM_s \to \mathcal{P}(W_s)$. For all $k \in NOM_s$, $V_s(k)$ is required to be a \textit{singleton} subset of $W_s$. This is the usual definition of valuations  on nominals in hybrid modal logic but, in this paper, we will simply denote $V_s(i)=w$ instead of $V_s(i)=\{w\}$ for $w\in W_s$ and $i\in NOM_s$. As customary in hybrid settings, Tarski-style assignment functions are introduced for the evaluation of variables:
given a $\bf\Sigma$-frame $\mathcal{F} = (W, (R_\sigma)_{\sigma \in \Sigma}, N^{\mathcal{F}})$, an assignment is an $S$-sorted function $(g_s)_{s \in S}$, $g_s : SVAR_s \to W_s$. We write $g'\rightsquigarrow^x g$ with the usual meaning of $g'$ being an $x$-variant of $g$. Note that, for any $s\in S$,  state symbols are evaluated in worlds  (elements of $W_s$), while  (modal)  propositional variables $p\in PROP_s$  are evaluated as arbitrary subsets of $W_s$.

Now we are ready to define the $S$-sorted  \textit{satisfaction relation}. Given a model $\mathcal{M} = (W, (R_\sigma)_{\sigma \in \Sigma}, N, V)$, an assignment function $g$ and $w \in W_s$, the relation 
  $\vDash=(\vDash^s)_{s\in S}$ is defined recursively for all $s \in S$ as in Table \ref{table:satisfaction} below. 
 The definition of validity in models and frames is standard.

Let $\Lambda$ be an $S$-sorted set of $\mathcal{H}_{\bf \Sigma}(@, \forall)$ formulas. Of particular importance to our considerations are the following two classes of models and frames: $Mod(\Lambda)$, the class of models in which every formula in $\Lambda$ is valid, and  $Fr(\Lambda)$,   the class of models based on frames where every formula in $\Lambda$ is valid. Clearly, $Fr(\Lambda) \subseteq Mod(\Lambda)$.

The Hilbert-style proof system for $\mathcal{H}_{\bf \Sigma}(@, \forall)$ is defined in Table \ref{table:axioms} below. Note that, for any $s\in S$ and $\varphi$ a formula of sort $s$ we denote by $\vdash^s\varphi$ the fact that $\varphi$ is provable. If $\Lambda$ is an $S$-sorted set of $\mathcal{H}_{\bf \Sigma}(@, \forall)$ formulas, then  $\mathcal{H}_{\bf \Sigma}(@, \forall) + \Lambda$ is the proof system obtained by extending $\mathcal{H}_{\bf \Sigma}(@, \forall)$ with the formulas of  $\Lambda$ as additional axioms, closed under the deduction rules from Table \ref{table:axioms}. We write $\vdash^s_{\Lambda} \varphi$ to denote that $\varphi$ is a theorem in the system 
$\mathcal{H}_{\bf \Sigma}(@, \forall) + \Lambda$. Axiomatic extensions are essential to the formalization of our applications from Section \ref{applications}.

{
\begin{table}[hbt!]
\centering
\footnotesize
\fbox{\parbox{355pt}{If $\mathcal{M} = (W, (R_\sigma)_{\sigma \in \Sigma}, N, V)$ is a model, $g$ is an assignment function, $s\in S$ and $w \in W_s$, then: 
    \begin{itemize}
        \item $\mathcal{M}, g, w \vDash^s p$ iff $w \in V_s(p)$, for $a \in PROP_s$;
\item  $\mathcal{M}, g, w \vDash^s i$ iff $w =V_s(i)$, for $i \in NOM_s$;
        \item $\mathcal{M}, g, w \vDash^s c$ iff $w = N^{\mathcal{F}}_s(c)$, for $c \in N_s$;
        \item $\mathcal{M}, g, w \vDash^s x$ iff $w = g_s(x)$, for $x \in SVAR_s$;
        \item $\mathcal{M}, g, w \vDash^s \neg \varphi$ iff $\mathcal{M}, g, w \nvDash^s \varphi$;
        \item $\mathcal{M}, g, w \vDash^s \varphi \lor \psi$ iff $\mathcal{M}, g, w \vDash^s \varphi$ or $\mathcal{M}, g, w \vDash^s \psi$;
        \item $\mathcal{M}, g, w \vDash^s \sigma(\varphi_{1},\dots,\varphi_{n})$ iff there exists $(w_1, \dots, w_n) \in W_{s_1} \times \dots \times W_{s_n}$ \\ \qquad such that $(w, w_1, \dots, w_n) \in R_\sigma$ and $\mathcal{M}, g, w_i \vDash^{s_i} \varphi_i$ for all $1 \leq i \leq n$;
        \item $\mathcal{M}, g, w \vDash^s @^s_k \varphi$ iff $\mathcal{M}, g, u \vDash^t \varphi$, where $k$ and $\varphi$ have sort $t$ and $V_t^N(k) = u$;
        \item $\mathcal{M}, g, w \vDash^s \forall x \varphi$ iff $\mathcal{M}, g', w \vDash^s \varphi$ for all $g'\rightsquigarrow^x g$.
    \end{itemize}}}
\caption{The satisfaction relation for the models of \textbf{$\mathcal{H}_{\bf \Sigma}(@, \forall)$}}
\label{table:satisfaction}
\end{table}
}

{
\begin{table}[hbt!]
\centering
\footnotesize
\begin{tabular}{|l|}
\hline
We assume that $s,t\in S$,  $\sigma\in \Sigma$ has an appropriate arity, and  $j,k\in NOM\cup N$, $x,y\in SVAR$,\\ $\varphi, \psi, \varphi_1,\dots, \varphi_n$ are formulas of appropriate sorts.\\
\textbf{The axioms of $\mathcal{H}_{\bf \Sigma}(@, \forall)$ are:}\\
\begin{tabular}{l l}
Prop & All propositional tautologies \\
K & $\vdash^s \sigma^{\Box} (\varphi_1,\dots, \varphi \to \psi, \dots,\varphi_n) \to \sigma^{\Box} (\varphi_1,\dots, \varphi, \dots,\varphi_n) \to \sigma^{\Box} (\varphi_1,\dots, \psi, \dots,\varphi_n)$ \\
Dual & $\vdash^s \sigma(\varphi_1, \dots, \varphi_n) \leftrightarrow \neg \sigma^{\Box}(\neg \varphi_1, \dots, \neg \varphi_n)$ 
\end{tabular}\\
\begin{tabular}{l l l l}
K@ & $\vdash^s @_j^s (\varphi_t \to \psi_t) \to (@_j^s \varphi \to @_j^s \psi)$ & SelfDual & $\vdash^s @_j^s \varphi_t \leftrightarrow \neg @_j^s \neg \varphi$\\
Agree & $\vdash^s @_k^s @_j^{s'} \varphi_t \leftrightarrow @_j^s\varphi_t$ &
Back & $\vdash^s \sigma(\dots,@_j^{s_i} \psi_t,\dots)_s \to @_j^s \psi$ \\
Intro & $\vdash^s j \to (\varphi_s \leftrightarrow @_j^s \varphi_s)$  & Ref & $\vdash^s @_j^s j$ 
\end{tabular}\\
\begin{tabular}{l l}
Q1 & $\vdash^s \forall x (\varphi \to \psi) \to (\varphi \to \forall x \psi)$, if $x$ does not occur free in $\varphi$ \\
Q2 & $\vdash^s \forall x \varphi \to \varphi[y / x] $, if $y$ is substitutable for $x$ in $\varphi$ \\
Barcan & $\vdash^s \forall x \sigma^{\Box}(\varphi_1,\dots, \varphi_i, \dots,\varphi_n ) \to \sigma^{\Box}(\varphi_1,\dots, \forall x \varphi_i, \dots,\varphi_n)$, \\ & \quad  if $x$ does not occur free in $\varphi_j$ for $j \neq i$ 
\end{tabular}\\
\begin{tabular}{l l l l l l}
Barcan@ & $\vdash^s \forall x @_j \varphi \to @_j \forall x \varphi$ & 
Nom & $\vdash^s @_k x \wedge @_j x \to @_k j$ & Name & $\vdash^s \exists x x$ 
\end{tabular}\\
\hline
\textbf{The deduction rules of $\mathcal{H}_{\bf \Sigma}(@, \forall)$ are:}\\
\begin{tabular}{l l}
MP & If $\vdash^s \varphi \to \psi$ and $\vdash^s \varphi$, then $\vdash^s \psi$ \\
UG & If $\vdash^{s_i}\varphi_i$, then $\vdash^s \sigma^{\Box}(\varphi_1,\dots,\varphi_i,\dots,\varphi_n)$ \\
BroadcastS & If $\vdash^s @_j^s \varphi_t$, then $\vdash^{s'} @_j^{s'} \varphi_t$ \\
Gen@ & If $\vdash^s \varphi$, then $\vdash^{s'} @_j^{s'} \varphi$ \\
Name@ & If $\vdash^s @_j^s \varphi$, then $\vdash^{s'} \varphi$, if $j \not\in N $ and $j$ does not occur in $\varphi$ \\
Paste & $\vdash^s @_j^s \sigma(\varphi_1,\dots , k, \dots,\varphi_n) \wedge @_k^{s} \varphi \to \psi$, then $\vdash^s @_j^s \sigma(\varphi_1,\dots, \varphi, \dots,\varphi_n) \to \psi$,\\ 
 & for $k \neq j$, $k \not\in N$ and $k$ does not occur in $\varphi$, $\psi$, or the $\dots$ formulas\\
Gen & If $\vdash^s \varphi$, then $\vdash^s \forall x \varphi$ \\
\end{tabular}\\
\hline
\end{tabular}
\caption{The provability relation for \textbf{$\mathcal{H}_{\bf \Sigma}(@, \forall)$}}
\label{table:axioms}
\end{table}
}

We further define the notions of \textit{semantic} and \textit{syntactic} consequence. In the sequel, we further assume that $s\in S$ and   $\Gamma$ be a set of formulas of sort $s \in S$. 
If $\mathcal{C}$ is a class of models,   we write $\Gamma \vDash^s_{\mathcal{C}} \varphi$ to denote that $\varphi$ is \textit{semantically entailed} by $\Gamma$ in all models from class  $\mathcal{C}$: for any model $\mathcal M$, for any assignment $g$ and for any $w\in W_s$, 
${\mathcal M},g,w\vDash^s \Gamma$ implies ${\mathcal M},g,w\vDash^s \varphi$. 
Note that our treatment focuses on the notion of \textit{local entailment}, as defined in \cite[1.5]{mod}. Recall that the \textit{local syntactic consequence}  $\Gamma \vdash^s_\Lambda \varphi$ is defined as follows: there exist formulas $\varphi_1 \dots \varphi_n \in \Gamma$, such that $\vdash^s_\Lambda (\varphi_1 \wedge \dots \wedge \varphi_n) \to \varphi$.  
 

Finally, we state the  \textit{soundness theorem} with respect to axiomatic extensions. In the following sections, we will mechanize its proof in Lean.

\begin{theorem}[Frame-soundness]\label{thsound}
    Let $\Lambda$ be an $S$-sorted set of formulas, and let $\Gamma$ be a set of formulas of sort $s \in S$. 
    Then $\mathcal{H}(@, \forall) + \Lambda$ is \textit{sound}, i.e. the following implication holds  
    for all $\varphi$ of sort $s$: $\Gamma \vdash^s_{\Lambda} \varphi \text{ implies } \Gamma \vDash^s_{Fr(\Lambda)} \varphi$.
\end{theorem}

\section{Representing the Syntax}\label{syntax}

A central goal of the formalization was to obtain an \textit{intrinsically sorted} representation of formulas, by clever use of Lean's dependent typing. In other words, we make no use of any "sort checking" function operating on syntax and defined separately from the syntax itself. Rather, any formula that can be constructed \textit{is guaranteed by definition} to be well-sorted, making further checks redundant. Since our end-goal is to provide an interface between programming language syntax and many-sorted signatures, intrinsic sorting ensures that \textit{programs are formulas}. More precisely, the programming language grammar is defined as a many-sorted signature; and, due to intrinsic sorting, valid parses in this grammar are guaranteed to be representable as hybrid formulas. We will show a few practical examples of how grammars define many-sorted signatures in section \ref{dsl}.

We employ an approach similar to the one laid out in \cite{gunther2018formalization}, by fixing a set \lstinline{S} of an arbitrary type \lstinline{α} to stand for \textit{sorts}, and letting all other families of symbols be dependent on \lstinline{S} (operators \lstinline{«Σ»}, constant nominals \lstinline{N}, nominals \lstinline{nom}, propositional and state variables \lstinline{prop, svar}). We show only the structure \lstinline{Signature α}, and note we also define a structure \lstinline{Symbols α}, made of a signature and the \lstinline{prop, nom, svar} S-indexed families of symbols described:

\begin{lstlisting}
structure Signature (α : Type u) where
  S    : Set α
  «Σ» : List S → S → Set α
  N    : S → Set α
\end{lstlisting}

Yet we diverge from \cite{gunther2018formalization} and other dependently typed formalizations of many-sorted languages, like \cite{depmatching}, by our treatment of \textit{modal operators}. In this respect, we developed what we call \textit{the list trick}, which we believe can be reused in other implementations of sorted languages in dependently-typed proof assistants. We describe it below.

\subsection{The List Trick}

As is usual, formulas are represented using an inductive type, each constructor corresponding to one kind of syntax: atomic symbols, logical connectives and operators, or modal applications. This type is parametrized by a variable \lstinline{symbs : Symbols α}, as defined earlier. Importantly, the type of formulas is further indexed by \textit{a list of sorts} in the many-sorted signature. Usually, we only care about singleton lists of sorts. For example, the \lstinline{at} constructor corresponds to the $@$ operator, and it ensures that you can take a nominal $k$ of sort $t$, a formula $\varphi$ of the same sort $t$, and you may obtain a formula $@_k^s \varphi$ of any arbitrary sort $s$:

\begin{lstlisting}
inductive FormL (symbs : Symbols α) : List symbs.signature.S → Type u
  | at   : symbs.nominal t → FormL symbs [t] → FormL symbs [s]
  -- ...other constructors...
\end{lstlisting}

Yet we also introduce an artificial syntax constructor, \lstinline{cons}, not present in the pen-and-paper definition of formulas. It is meant to build \textit{lists} of formulas indexed by \textit{non-singleton} lists of sorts, by \textit{consing} (in the usual functional programming sense) a single sort formula onto a list of other sorted formulas:

\begin{lstlisting}
  | cons : FormL symbs [s₁] → FormL symbs (s₂ :: t) → FormL symbs (s₁ :: s₂ :: t)
\end{lstlisting}

In our setting, this trick is a very natural choice. Recall that a modal operator in our many-sorted settings has \textit{domain} sorts and a \textit{range} sort: when an operator is applied to a list of formulas that are well sorted according to its domain, it yields a formula in its range sort. Using the \lstinline{cons} constructor, we gain the ability to construct such heterogeneously sorted argument lists seamlessly. The sorts of the arguments are kept in the index of the \lstinline{FormL} type. Therefore, the condition that the arguments are well sorted reduces to their index being identical to the domain of the operator. This is a verification that the Lean type checker itself can make, giving us intrinsic sorting. Our modal application constructor is, therefore, introduced like this:

\begin{lstlisting}
  | appl : symbs.signature.Σ (h :: t) s → FormL symbs (h :: t) → FormL symbs [s]
\end{lstlisting}

By contrast, the approach taken in \cite{gunther2018formalization, depmatching} is to introduce the type of \textit{heterogeneous lists}, i.e., lists of terms of possibly different types. In our case, this would amount to the following alternative formalization (see also \cite{Chlipala2013-ot} for more on \lstinline{HList}):
\begin{lstlisting}
inductive HList {α : Type v} (f : α → Type u) : List α → Type (max u v)
  | nil  : HList f []
  | cons : f x → HList f xs → HList f (x :: xs)
inductive Form {symbs : Symbols α} : symbs.signature.S → Type u
  | appl : symbs.signature.Σ (h :: t) s → HList Form (h :: t) → Form s
\end{lstlisting}

We think the \textit{list trick} is a more appropriate formalization choice in contexts, like ours, where one will be prone to intensively recurse and perform induction over formulas. The reason is that it is easier on the proof assistant's compiler: it requires neither nested, nor mutual inductive types. Such inductives currently lack convenient support for the \lstinline{induction} tactic in Lean, and they create additional proof burden to show termination on recursive definitions.

The use of heterogeneous lists we showed above involved a nested inductive (\lstinline{HList} mentioned within \lstinline{Form}). Indeed, the authors of \cite{depmatching} acknowledge that "[an] issue arises [...] the argument
list of the application constructor. [...] This makes recursive definitions and inductive proofs over the type difficult, if not impossible". Using the \textit{list trick}, we obtain recursion and induction for free, at no expense to expressivity.

We still use a form of heterogeneous lists, disguised as heterogeneous \textit{products}, through a type we called \lstinline{WProd}. We will introduce this type in Section \ref{semantics}, where we go over the semantics.

Besides the list trick, all other constructors to \lstinline{FormL} are standard. We defined list-like notation for the \lstinline{cons} constructor: \lstinline{(φ, ψ)} denotes \lstinline{FormL.cons φ ψ}. Pen-and-paper formulas then correspond to singleton \lstinline{FormL}'s:

\begin{lstlisting}
def Form (symbs : Symbols α) (s : symbs.signature.S) := FormL symbs [s]
\end{lstlisting}

We note that we made the choice to represent variables concretely, using their names; as opposed to other approaches common in the literature, such as de Bruijn indices:

\begin{lstlisting}
  | svar : symbs.svar s → FormL symbs [s]
  | bind : symbs.svar t → FormL symbs [s] → FormL symbs [s]
\end{lstlisting}

The intention was to keep the representation of syntax as close to its pen-and-paper counterpart as possible. Since flavors of hybrid logic with different binders can be defined (e.g., with an additional $\downarrow$ operator, or where the $@$ operator can also bind variables), our representation of variables as concrete allows us to quickly iterate through these different versions without requiring additional engineering effort to make them compatible with our choice of variable naming. The point was not so much to define a \textit{static formal artifact} of the logic, but rather to provide a \textit{research tool} that the pen-and-paper logician can seamlessly migrate to and from when conducting research, with little representational overhead. We therefore also provide the usual pen-and-paper definitions for substitution, along with a decidable predicate \lstinline{free_for} that checks whether a variable is substitutable for another in a given formula (as per Table \ref{table:axioms}, where these notions are also employed).

\subsection{Mechanizing the Proof System}\label{proofsys}

The generality of this system forces the pen-and-paper axiomatization (Table \ref{table:axioms}) to be somewhat non-rigorous, in particular with respect to axioms $K$, $Back$, $Barcan$, and the $UG$ and $Paste$ rules. Since modal operators are allowed to have arbitrary arities, notice that the symbol "$\dots$" is used to refer to \textit{some} argument to an operator $\sigma$. Syntactically, the point is to perform \textit{substitution} upon this argument.
It is something specific to this polyadic framework that one is required to refer to operator arguments generically, and perform substitution upon them.
While the meaning of "$\dots$" is evident to the mathematical reader, it becomes a genuine question of engineering how this mechanism of reference and substitution should be expressed in Lean. To this end, we introduce the notion of \textit{contexts}\footnote{Note that our notion of \textit{context} is different from "a term with one hole", which is the usual meaning.}, and \textit{context substitution}.

\noindent{\bf Contexts.} A term of type \lstinline{Context φ ψ} can be seen as a \textit{pointer} to the occurrence of formula \lstinline{φ} within the formula (list) \lstinline{ψ}. Its definition is given below. It is rather similar to the \lstinline{member} type introduced in \cite{Chlipala2013-ot} with respect to \lstinline{HList}s; and could also be seen as a zipper \cite{huet1997zipper} on the \lstinline{FormL} data structure.

\begin{lstlisting}
inductive Context {symbs : Symbols α} {s : symbs.signature.S} (φ : Form symbs s) : FormL symbs sorts → Type u
  | refl : Context φ φ
  | head : Context φ (φ, ψ)
  | tail : Context φ ψ → Context φ (χ, ψ)
\end{lstlisting}

Notice that we do not define an inductive \textit{proposition}, but rather an inductive \textit{type}. That is, the existence of a term \lstinline{C : Context φ ψ} signifies not only \textit{the fact} that formula \lstinline{φ} occurs within \lstinline{ψ}; but it also carries the (unique) constructive description of \textit{where} said occurrence is to be found. We exemplify this below, where we list all possible contexts of some formula in the list \lstinline{(φ, ψ, φ)}.

\begin{lstlisting}
variable (φ ψ : Form symbs s)
example : φ.Context (φ, ψ, φ) := Context.head
\end{lstlisting}
\vspace{-1em}
\begin{lstlisting}
example : ψ.Context (φ, ψ, φ) := Context.tail Context.head
example : φ.Context (φ, ψ, φ) := Context.tail <| Context.tail Context.refl
\end{lstlisting}

\noindent Crucially, in the example above, the two different occurrences of \lstinline{φ} within \lstinline{(φ, ψ, φ)} correspond to two definitionally \textit{different} \lstinline{φ.Context (φ, ψ, φ)} terms.
Contexts, then, amount to a rigorous mechanism of specifying a formula within "$\dots$" notation. What is still required is to define the operation of \textit{substituting} said formula with a different one. We use the notation \lstinline{C[φ]} for the term \lstinline{subst C φ}. For brevity, we only show its type below; it takes a context \lstinline|C| and a new formula \lstinline|φ|, and returns the formula (list) obtained by replacing the argument that \lstinline|C| points to with \lstinline|φ|.

\begin{lstlisting}
def subst : Context φ ψ → Form sig s → FormL sig sorts
\end{lstlisting}

With this machinery in place, consider our Lean characterization of axiom K. We provide below the respective constructor of the \lstinline|Proof| type:
\begin{lstlisting}
  | k φ ψ χ (C : (φ ⟶ ψ).Context χ):
        Proof Λ s (ℋ⟨σ⟩ᵈ χ ⟶ (ℋ⟨σ⟩ᵈ C[φ] ⟶ ℋ⟨σ⟩ᵈ C[ψ]))
\end{lstlisting}

We express that \lstinline|φ ⟶ ψ| must occur somewhere among the arguments to \lstinline|ℋ⟨σ⟩ᵈ| by requiring a \lstinline|C : (φ ⟶ ψ).Context χ| term. We successively substitute that particular occurrence by \lstinline{φ} and \lstinline{ψ}, respectively, by applying the context substitutions \lstinline{C[φ]} and \lstinline{C[ψ]}.

\medskip
\noindent{\bf Axiomatic Extensions}\label{extensions}
An important design question was that of axiomatic extensions. We defined sets of axioms simply as $S$-sorted sets of formula:

\begin{lstlisting}
def AxiomSet (symbs : Symbols α) := (s: symbs.signature.S) → Set (Form symbs s)
\end{lstlisting}

Our \lstinline|Proof| type is parametric on an axiomatic extension \lstinline|Λ : AxiomSet symbs|, which may also be empty. We ensure that all theorems generated by the axioms extended with \lstinline|Λ| are \lstinline|Proof| terms by adding an additional \lstinline|ax| constructor:

\begin{lstlisting}
  | ax    : (φ : Λ s) → Proof Λ s φ
\end{lstlisting}

This guarantees that not only formulas from \lstinline|Λ| are taken as theorems, but also their closure under the application of any number of deduction rules, possibly also using axioms from the base system. 
While simple to implement, this approach is also very powerful, allowing both finitary and infinitary axiomatic extensions. In particular, in Section \ref{operational} we show how the proof system can be extended infinitarily by a number of \textit{finitely specified} axiom \textit{schemas}.

\medskip
\noindent{\bf The Proof System, Mechanized.} Since the full proof system led to a large Lean definition, below we display only an excerpt of the \lstinline|Proof| inductive type, which suffices to showcase the hybrid nature of the language, and our design choices with respect to contexts.

\begin{lstlisting}
inductive Proof {symbs : Symbols α} (Λ : AxiomSet symbs) : (s : symbs.signature.S) → Form symbs s → Type u
  | agree j k φ  : Proof Λ s (ℋ@k (ℋ@j φ) ←→ ℋ@j φ)
  | q1 x φ ψ (_: φ.occurs_free x = false): Proof Λ s (ℋ∀x (φ ⟶ ψ) ⟶ (φ ⟶ ℋ∀x ψ))
  | q2_var (x y: symbs.svarType s') φ: φ.free_for y x → Proof Λ s ((ℋ∀x φ) ⟶ φ[y // x])
  | name x : Proof Λ s (ℋ∃x (ℋVar x))
  | barcan x (φ : Form symbs s) (σ : symbs.signature.Σ _ s) (C : φ.Context ψ)
      (h : C.all_else_not_free x):
            Proof Λ s (ℋ∀x (ℋ⟨σ⟩ᵈ ψ) ⟶ (ℋ⟨σ⟩ᵈ C[ℋ∀x φ]))
  | paste {k : symbs.nomType s₂} {φ : Form symbs s₂} (C : (ℋNom k).Context χ):
          .nom k ≠ₛ j → ψ.occurs k = false → C[φ].occurs k = false →
\end{lstlisting}
\vspace{-1em}
\begin{lstlisting}
          Proof Λ s₁ (ℋ@j (ℋ⟨σ⟩ χ) ⋀ ℋ@k φ ⟶ ψ) → Proof Λ s₁ ((ℋ@j (ℋ⟨σ⟩ C[φ]) ⟶ ψ))
\end{lstlisting}

In the following section, we will describe our treatment of the Kripke semantics for this logic.

\section{Representing the Semantics. Models and Soundness}\label{semantics}
Let us begin by describing our approach regarding the $S$-sorted sets of worlds. Our frames map each sort to some type in a universe \lstinline|u|:

\begin{lstlisting}
structure Frame (signature : Signature α) where
  W  : signature.S → Type u
  -- ..other fields...
\end{lstlisting}

This represents the fact that \textit{each term of type} \lstinline|W s| is \textit{a world of sort} \lstinline|s|. Recall, now, that each $S$-sorted operator $\sigma$ is interpreted as a subset $R_\sigma$ of the cartesian product of correspondingly sorted worlds. In order to define $R_\sigma$, we needed an auxiliary function that takes a list of sorts ($\sigma$'s domain and range) and $W$, and returns the \textit{product type} of the list of sorts mapped through $W$:\footnote{In order to define a total function, the edge case of \texttt{sorts} being empty must be considered, but note that all operators have at least \textit{one} sort, namely their \texttt{range}. This is a basic assumption of signatures both in the pen-and-paper version, and in our Lean formalization of it.}

\begin{lstlisting}
def WProd {sig : Signature α} (W : sig.S → Type u) : List (sig.S) → Type u
  | []      => PEmpty
  | [s]     => W s
  | s :: sorts  => W s × WProd W sorts
\end{lstlisting}

A term of type \lstinline|ws : WProd W sorts|, where \lstinline|sorts = [s, s₁, ..., sₙ]| corresponds to an element of the cartesian product $W_s \times W_{s_1} \times \dots \times W_{s_n}$. Note that \lstinline{WProd W sorts} is definitionally equal to a product type when \lstinline{sorts} has at least two elements. This allows us to destruct such a \lstinline{WProd} into its first and second components, a fact that we exploit in the satisfaction relation for formula lists (more on this below).

Having the definition of \lstinline{WProd} on hand, the definitions of models and frames follow cleanly, and we list them in full below (certain restrictions we impose on \lstinline{W} notwithstanding). The accessibility relation of an operator is introduced as a \textit{set of \lstinline{WProd}'s}, denoting all the sorted worlds that stand together in the respective relation. We define the valuation functions \lstinline{Nm}, \lstinline{Vₚ} and \lstinline{Vₙ}, which specify the worlds where constant nominals, propositional symbols and non-constant nominals are satisfied, respectively:
\begin{lstlisting}
structure Frame (signature : Signature α) where
  W  : signature.S → Type u
  R  : signature.Σ dom range → Set (WProd W (range :: dom))
  Nm : {s : signature.S} → signature.N s → W s
structure Model (symbs : Symbols α) where
  Fr  : Frame symbs.signature
  Vₚ  : symbs.prop s → Set (Fr.W s)
  Vₙ  : symbs.nom s → Fr.W s
\end{lstlisting}

Following the slight generalization induced by the \textit{list trick} described in Section \ref{syntax}, the satisfaction relation is also slightly generalized in the following sense. We define the \textit{simultaneous satisfaction} of a formula list $\varphi_1, \dots, \varphi_n$ at a product of worlds $W_1 \times \dots \times W_n$, as the satisfaction of a \lstinline{FormL} at a \lstinline{WProd}, thus:

\begin{lstlisting}
def Sat (M : Model symbs) (g : Assignment M) (w : WProd M.Fr.W sorts) : FormL symbs sorts → Prop
  | .cons φ ψs     => Sat M g w.1 φ ∧ Sat M g w.2 ψs
\end{lstlisting}

Once again, this list trick eases our treatment of modal applications. An operator \lstinline{σ} applied to a formula \textit{list} \lstinline{arg} is satisfied at world \lstinline{w}, if and only if, there exists a product \lstinline{w' : WProd} that is accessible from \lstinline{w}, such that \lstinline{arg} is simultaneously satisfied at \lstinline{w'}. Formally:
\begin{lstlisting}
  | .appl σ arg    => ∃ w', Sat M g w' arg ∧ ⟨w, w'⟩ ∈ M.Fr.R σ
\end{lstlisting}

The rest of the satisfaction relation is standard and we will not list it in full.
We use the notation \lstinline{⟨M, g, w⟩ ⊨ φ} for \lstinline{Sat M g w φ}. We further defined entailment, validity in a class of models, validity in a class of frames, and validity in the class of frames determined by the axiom set \lstinline{Λ}. We denote the latter as \lstinline{⊨Fr(Λ) φ}.

\subsection{Mechanizing Soundness}

Using these definitions of syntax, proof system and semantics, we were successful in writing a complete, sorry-free formalization of the following soundness statements, the second of which being an immediate consequence of the first:
\begin{lstlisting}
theorem Soundness {Λ : AxiomSet symbs} : ⊢(Λ, s) φ → ⊨Fr(Λ) φ := by
theorem StrongSoundness {Λ : AxiomSet symbs} : Γ ⊢(Λ) φ → Γ ⊨Fr(Λ) φ := by
\end{lstlisting}

In the process of formalizing soundness, we identified several errors in the original presentation of the proof system given in \cite{tableaux}, namely with respect to axiom Barcan, and rules Name@ and Paste, which were missing some of the syntactic side conditions that are now shown in Table \ref{table:axioms}. We were able to construct a countermodel in Lean to the original version of Barcan, using the formalization presented in this paper. The proof system we presented in Section \ref{prel} corrects these issues, ensuring soundness. With this out of the way, the actual soundness proof is standard and follows by induction on \lstinline{Proof} terms. Consider the proof for Name@:
\begin{lstlisting}
| @nameAt s₁ s₂ j φ noccφ _ ih =>
    intro M g w
    let M' := ⟨M.1.v_variant j w, Set.Elem.v_variant_modelclass_inv Λ M j w⟩
    let g' : Assignment M'.1 := g.v_variant j w
    let w' := (M'.1.Fr.WNonEmpty s₁).default
    specialize ih M' g' w'
    simp only [Sat.at, M', v_variant_valuation] at ih
    rw [v_variant_agreement noccφ w]
    exact ih
\end{lstlisting}

It proceeds by defining the $M'$ model as a valuation variant, while also introducing $g'$ and $w'$ definitions (identical to $g$ and $w$, but made to match $M'$'s set of worlds). It then specializes the inductive hypothesis (namely, $\vDash^s_{Fr(\Lambda)} @_j \varphi$) to $M'$, $g'$, and $w'$, and after simplification with the satisfaction clauses of the $@$ operator, obtains $M', g', V'(j) \vDash^s \varphi$. Finally, using the variant agreement lemma, whose statement is shown below, it rewrites this to $M, g, w \vDash^s \varphi$, which is exactly the statement we intended to prove.

\begin{lstlisting}
lemma v_variant_agreement {ws : WProd M.Fr.W sorts} {g : Assignment M} {j : symbs.nominal s} (h : φ.occurs j = false) (w : M.Fr.W s) :
  (⟨M, g, ws⟩ ⊨ φ) ↔ ⟨M.v_variant j w, g, ws⟩ ⊨ φ :=
\end{lstlisting}

It states that, for any formula \lstinline{φ}, we are free to alter the valuation of a nominal \lstinline{j} in a model \lstinline{M} without affecting the satisfaction of \lstinline{φ}; given that \lstinline{j} does not occur in \lstinline{φ}.

\section{Applications}\label{applications}
We showcase the possible applications of our system with three examples.
\begin{itemize}
\item The \textit{SMC language} from \cite{plotkin, tableaux}, for which we are able to perform  Hoare-like verification for simple programs.
\item \textit{Logics for analysing security protocols}: the BAN logic, developed by Burrows, Abadi, and Needham  \cite{ban} and an epistemic logic with actions, adapted from \cite{ramics}. We analyze two concrete protocols:  One Sided Secrecy, and the Needham-Schroeder Public Key Protocol (which is a standard example for any formal method approach of this domain).
\item \textit{The S5 modal logic}. We claim that the system we defined is modular enough to serve as a general-purpose \textit{modal logic toolbox}, applicable even to cases where the full expressive power of the base system is unneeded. We axiomatize S5 in the strictly \textit{modal} fragment of the language.
\end{itemize}

\subsection{A DSL for Language Definitions}\label{dsl}
We provide a \textit{DSL for defining many-sorted hybrid signatures} in familiar BNF-like notation. We start by briefly explaining the syntax and the elaboration process. All applications we will show in the subsequent sections will make use of our custom domain-specific syntax. 

Signature definitions are introduced via the \lstinline{hybrid_def} keyword. The definition may span over multiple lines, each of which declares the symbols corresponding to some sort. The following is a simple example declaring three sorts:

\begin{lstlisting}
hybrid_def Ex :=
  sort One   ::= builtin Nat | "op₁"(One, One)
  sort Two   ::= nom₁ | nom₂ | "op₂"(Two)
  sort Three ::= subsort One
\end{lstlisting}

The \lstinline|builtin| keyword declares a new constant nominal of sort \lstinline|One| for each natural number expressed as a Lean term of type \lstinline|Nat|. Additionally, \lstinline|nom₁| and \lstinline|nom₂| are declared as constant nominals of sort \lstinline|Two|. Two modal operators are declared. First, we have \lstinline|op₁|, taking two formulas of sort \lstinline|One| and returning a formula of sort \lstinline|One|. Second, there is \lstinline|op₂|, which takes a formula of sort \lstinline|Two| and returns another formula of sort \lstinline|Two|. 
That is, even though the sorted signatures of this logic do not support subsorting, the declaration of \lstinline|sort Three ::= subsort One| creates an operator \lstinline|one2Three| which maps formulas of sort \lstinline|One| onto formulas of sort \lstinline|Three|. Internally, this definition is elaborated as a \lstinline|Symbols String| declaration, along with a \lstinline|Signature String| declaration it uses internally: \lstinline{#check Ex -- Ex : Symbols String}

Each of the constituting fields of the \lstinline{Symbols} structure must  be filled during elaboration. The following auxiliary declarations are also created and their names are accessible:
\begin{lstlisting}
#check Ex.Sig   -- Ex.Sig : Signature String
#check Ex.Sorts -- Ex.Sorts : Set String
#check Ex.Ops   -- Ex.Ops (domain : List ↑Ex.Sorts) (range : ↑Ex.Sorts) : Set String
#check Ex.CtNoms -- Ex.CtNoms (s : ↑Ex.Sorts) : Set String
\end{lstlisting}

The mechanism of elaborating these $S$-sorted sets is kept as simple as possible. \lstinline{Ex.Ops} and \lstinline{Ex.CtNoms} are both defined as actual \textit{functions}, which decide membership in their respective sorted sets by checking boolean equality with the signature of each symbol. To exemplify this, we show the full elaborated definition of \lstinline{Ex.Ops}:

\begin{lstlisting}
def Ex.Ops : List ↑Ex.Sorts → ↑Ex.Sorts → Set String :=
\end{lstlisting}
\vspace{-1em}
\begin{lstlisting}
  fun domain range => setOf ((fun domain range operator =>
    (bif domain == [Ex.One, Ex.One] && (range == Ex.One && operator == "op₁One_One_One") then true
      else
        bif domain == [Ex.Two] && (range == Ex.Two && operator == "op₂Two_Two") then true
        else
          bif domain == [Ex.One] && (range == Ex.Three && operator == "one2Three") then true else false) =
      true) domain range)
\end{lstlisting}

While such definitions could be given by hand, thanks to DSL syntax the user is spared from the tedious and error-prone process of writing them.

\subsection{Operational Semantics: SMC Machine}\label{operational}
We begin by showing how the syntax and semantics of an imperative language can be defined within this framework. We focus on the SMC machine, introduced by Plotkin in \cite{plotkin}. Its syntax is similar to the textbook IMP language \cite{winskel1993formal}, while its small-step operational semantics makes explicit use of a lower-level \textit{stack-memory-control} model. The hybrid treatment of this language originates in \cite{tableaux}, our main target of Lean formalization. We specify the syntax of the SMC machine in Lean as a \textit{many-sorted hybrid logic signature}; we provide a corresponding \textit{axiomatic extension} to model its operational semantics; and we show how Hoare-like reasoning can be performed inside this formalization.

{\bf Describing the Grammar.} Using our DSL, the grammar of the SMC machine was succinctly described, and automatically elaborated into a hybrid signature definition. For brevity, below we provide an excerpt. The syntax between square brackets \lstinline{[...]} introduces a user-facing name for the defined operator (e.g., once elaborated, \lstinline{PlusNat} is the name of a Lean declaration of a binary operation on sort \lstinline{Nat}).

\begin{lstlisting}
hybrid_def SMC :=
\end{lstlisting}
\vspace{-4mm}
\begin{lstlisting}
    sort Nat  ::= builtin Nat | "_+_"(Nat, Nat)     [PlusNat]
    sort Stmt ::= skip
                | "_:=_"(Var, AExp)
                | "if_then_else_"(BExp, Stmt, Stmt) [IteStmt]
                | "while_do_"(BExp, Stmt)           [WhileStmt]
                | "_;_"(Stmt, Stmt)                 [SeqStmt]
    sort CtrlStack ::=  "c"(Stmt)                   [cStmt]
                | "_∪_"(CtrlStack, CtrlStack)       [PDLUnion]
                | "_;_"(CtrlStack, CtrlStack)       [PDLSeq]
                | "*"(CtrlStack)                    [PDLStar]
    sort Config ::= "<_,_>"(ValStack, Mem)          [mkConfig]
                | "[_]_"(CtrlStack, Config)         [PDLOp]
\end{lstlisting}

The code above elaborates to the declaration of a many-sorted signature \lstinline{SMC}. Here and elsewhere, we abbreviate the type \lstinline{Form SMC}, that of \textit{formulas in the signature} \lstinline{SMC}, by \lstinline{SMCForm}; and similarly for signatures introduced in later sections.

Consider the \lstinline{sort Stmt ::= if_then_else_(BExp, Stmt, Stmt)} production. It declares \lstinline{if_then_else_} as a polyadic modal operator, taking as arguments formulas of sorts \lstinline|BExp| (the branch condition), \lstinline|Stmt| (the "true" branch) and \lstinline|Stmt| (the "false" branch), and returns a \lstinline|Stmt| (the if-then-else statement itself).

Yet it would obviously be burdensome to be forced to write statements in a programming language directly as raw modal logic operator applications. This is not the path we have chosen. Rather, we define additional syntax macros that translate high-level programming-like syntax to our low-level logic, enabling us to finally write programs (i.e., \textit{sorted hybrid formulas}) exactly as one could provide as input to a regular compiler:

\begin{lstlisting}
def incrementMax (x y aux : SMCForm Var): SMCForm Stmt :=
  if (x <= ++y) then
    aux ::= x;
    x   ::= y;
    y   ::= aux
  endif
\end{lstlisting}

The type of this declaration indicates that it defines a \textit{hybrid formula of sort} \lstinline|Stmt|, in the signature of \lstinline|SMC|. As explained in Section \ref{syntax}, note that \textit{programs are (well sorted) formulas}. Thanks to our intrinsically sorted approach, Lean's type checker guarantees that any valid formula of sort \lstinline{Stmt} is a valid parse in SMC's grammar -- at no additional cost to the user other than writing the signature definition and its syntax macros.

\medskip
{\bf An Embedding of Hoare Logic. Towards Symbolic Execution.} Syntax is not all: we care about semantics. As explained in Section \ref{extensions}, one may extend the base proof system by providing an $S$-sorted set \lstinline|Λ| of additional axioms. Since SMC's operational semantics are \textit{not} finitely axiomatizable, but are introduced as axiom \textit{schemas}, it is another genuine question of proof engineering how SMC's infinite extension \lstinline|Λ| should be formally specified.

We believe Lean's type theory provides an elegant solution to this: inductive types. To characterize some \lstinline|Λ| by a finite number of schemas, one would start by defining an inductive family parametrized by formulas:

\begin{lstlisting}
inductive Axiom : {s : Sorts} → SMCForm s → Type
\end{lstlisting}

It's not the type \lstinline|Axiom| that we primarily care about; it is, rather, \textit{its indices}. We declare axiom schemas as constructors for the \lstinline|Axiom| type, an example being:

\begin{lstlisting}
| APlusNat {n₁ n₂ : ℕ}: Axiom ((n₁ +Nat n₂) ←→ (n₁ + n₂))
\end{lstlisting}

This schema constrains SMC's internal addition operator \lstinline|+Nat| to act in accordance to its model, namely mathematical addition on natural numbers. (As one might expect, this term could be type checked only after a number of coercions were defined.)

What we obtain by this constructor is that \textit{for each} \lstinline|n₁ n₂ : ℕ|, a term of type \lstinline|Axiom ((n₁ +Nat n₂) ←→ (n₁ + n₂))| is defined. Hence, this constructor acts as an axiom \textit{schema}, defining an infinite number of \lstinline|Axiom φ| terms, for varying \lstinline|φ|.  So we take \lstinline|Λ| precisely as the $S$-sorted set of formulas \lstinline|φ| for which a term of type \lstinline|Axiom φ| exists:

\begin{lstlisting}
def SMCΛ : AxiomSet SMC := λ _ => { φ | Nonempty (Axiom φ) }
\end{lstlisting}

The proof systems we obtain by extensions of this kind with \lstinline|Λ| are well suited for reasoning. In general we wrote helper definitions for easy access to the axioms, an example being:
\begin{lstlisting}
def aid : SMCProof _ (⟨vs, set(mem, x, n)⟩ ⟶ [c(x)] ⟨n ⬝ vs, set(mem, x, n)⟩) :=
    Proof.ax ⟨_, Nonempty.intro Axiom.Aid⟩
\end{lstlisting}

Let us return to the \lstinline|incrementMax| program we showed earlier. We make the following claim: if \lstinline|x| is initialized to 0 and \lstinline|y| is initialized to 2, then after the execution of \lstinline|incrementMax|, \lstinline|x| will have value 3 and \lstinline|y| will have value 0 (assuming all variables are distinct). Formally, this amounts to the following statement:

\begin{lstlisting}
def ifCorr (neq1 : x ≠ y) (neq2 : y ≠ aux) (neq3 : x ≠ aux): SMCProof _
      (⟨vs, set(set(mem, x, (0:ℕ)), y, (2:ℕ))⟩ ⟶
        [c(incrementMax x y aux)] ⟨vs, set(set(mem, x, (3:ℕ)), y, (0:ℕ))⟩) :=
\end{lstlisting}

Note, in particular, that the correctness statement above accounts for the side-effecting branch condition expression \lstinline|x <= ++y|. To perform Hoare-like verification, we minimally require a structural rule to help us reason about if-then-else statements, with the additional property of allowing branch conditions to side-effect. Following the mathematical exposition in \cite{tableaux}, we proved a number of such derived rules, with their respective derivations fully formalized in Lean. We exemplify by the rule of conditional below. Note the additional hypothesis \lstinline{h1} which allows the evaluation of the boolean expression \lstinline{b} to mutate memory:

\begin{lstlisting}
def conditional {b : SMCForm BExp}
\end{lstlisting}
\vspace{-1em}
\begin{lstlisting}
    (h1 : SMCProof _ (φ ⟶ [c(b)] ⟨B ⬝ vs, mem⟩))
    (h2 : SMCProof _ (⟨vs, mem⟩ ⋀ ℋ@ true B ⟶ [c(s₁)] χ))
    (h3 : SMCProof _ (⟨vs, mem⟩ ⋀ ℋ@ false B ⟶ [c(s₂)] χ)):
  SMCProof _ (φ ⟶ [c(if b then s₁ else s₂ endif)] χ) :=
\end{lstlisting}

With these derived rules on hand, constructing the proof \lstinline{ifCorr} becomes a matter of chaining \lstinline{SMCΛ} axioms with Hoare-like structural rules. The underlying modal embedding is once again abstracted: whoever engineers the proof does \textit{not} have to worry about using the low-level proof system from Section \ref{proofsys} directly. Further, proofs can be compositionally engineered, by first proving lemmas about subprograms, as we did for example by initially proving the correctness of the inner variable swap:

\begin{lstlisting}
def swapCorrect (neq1 : y ≠ x) (neq2 : y ≠ aux) (neq3 : x ≠ aux) : SMCProof _
    (⟨vs, set(set(mem, y, yn), x, xn)⟩ ⟶ [c(swapPgm x y aux)] ⟨vs, set(set(set(mem, x, yn), aux, xn), y, xn)⟩) :=
\end{lstlisting}

The work we display here with respect to Hoare verification focuses solely on concrete execution, as displayed in the examples above. A structural rule is proved in \cite{tableaux} under the name "Rule of Iteration", which would allow one to perform symbolic execution by proving loop invariants. This is currently an active area of proof engineering and still work-in-progress.

\subsection{Verification of Security Protocols}

We proceed by showcasing different logics for security protocols that can be embedded inside our hybrid modal machinery. We formalize BAN Logic \cite{ban}, capturing belief modalities, freshness, jurisdiction and message-meaning rules, and a hybrid modal logic for protocols \cite{ramics}, integrating dynamic actions (send and receive) with epistemic modalities, enabling reasoning about explicit knowledge and belief across protocols states. 

\medskip
{\bf An Embedding of BAN Logic.} A fragment of BAN formulas is provided using our DSL, as follows (for brevity, we omit some productions from the presentation).
\begin{lstlisting}
hybrid_def BAN :=
  sort Message ::= "↔"(Agent, Agent, Message) [shareKey]
      | "pk"(Agent)  | "sk"(Agent)
      | "enc"(Message, Message) [encrypt] | "pair"(Message, Message)
\end{lstlisting}
\vspace{-1em}
\begin{lstlisting}
  sort Formula ::= "|≡"(Agent, Formula) [believes]
      | "ι"(Message) | "|~"(Agent, Formula) [oncesaid]
      | "◁"(Agent, Formula) [sees] | "#"(Message) [nonce]
      | "|=>"(Agent, Formula)         [jurisdiction]
\end{lstlisting}

BAN deduction rules are encoded in the extension of the deduction system, and we provide one as example: 
\begin{lstlisting}
inductive Axiom : {s : Sorts} → BANForm s → Type
  | MMSK {i j : BANForm Agent} {m k : BANForm Message}
    : Axiom $ (i |≡ ι shareKeyOp i j k) ⟶
\end{lstlisting}
\vspace{-1em}
\begin{lstlisting}
        (i ◁ ι ⦃ m ⦄k) ⟶ (i |≡ j |∼ ι m)
\end{lstlisting}

This rule states that if an agent knows that it shares a key with another one, and it sees (receives) a message encrypted with that key, then it may assume that the message was indeed sent by the agent with whom the key is shared. 

By embedding the BAN rules into our formal framework, we obtain a mechanism for verifying authentication properties of security protocols. As an example, we prove that in the Needham-Schroeder public-key (NSPK) protocol, agent $B$ can derive that agent $A$ has received the nonce $n_B$ and believes it to be fresh, hence that it was sent within the current protocol session: 
\begin{lstlisting}
def NSPK_mutual_authentication (p₃ : BANProof _ $ B ◁ ι ⦃ nB ⦄sk(A))
  (a₁ : BANProof _ $ B |≡ ι pk(A)) (a₂ : BANProof _ $ B |≡ #(nB)):
    BANProof _ (B |≡ (A |≡ ι nB)) := by
  let h₁ : BANProof _ _ := Proof.ax ⟨_, Nonempty.intro $ Axiom.MMPK⟩
  let h₂ : BANProof _ $ (B ◁ ι ⦃ nB ⦄sk(A)) ⟶ (B |≡ (A |∼ ι nB)) := Proof.mp h₁ a₁
  let h₃ : BANProof _ $ B |≡ (A |∼ ι nB) := Proof.mp h₂ p₃
  exact NV h₃ a₂
\end{lstlisting}

{\bf An Embedding of Hybrid Logic for Security Protocols.} We refer to \cite{ramics} for a specific logic used in specifying and verifying security protocols, that can be implemented as a particular system of the current work, without any changes needed in representation of formulas. We provide only a fragment of the DSL used in our development: 
\begin{lstlisting}
hybrid_def Protocols :=
  sort Msg     ::= "⦃_⦄_"(Msg, Msg)           [encryption]
  sort Act     ::= "send"(Agent, Agent, Msg)  [send]
                 | "recv"(Agent, Msg)         [recv]
  sort StNom   ::= "◁"(Agent, Msg) [explicit] | "⊔"(StNom, StNom) [comp]
  sort Prot    ::= "B"(Agent, Prot) [believe] | "⟪_⟫"(StNom) [config]
                 | "[_]_"(Act, Prot) [action] | "X"(Agent, Msg) [explicitKnowledge]
\end{lstlisting}

Our goal  is to analyze a property for the  very simple protocol $OSS$ (one-sided secrecy):  $i \longrightarrow r: \{ i, n \}_{pk(i)}$, where $i$ and $r$ are two agents, $n$ is a message, and $pk(i)$ is the public key of the agent $i$. 
We formalize this protocol using the following two axioms: 
\begin{lstlisting}
| OSS₁ : Axiom $ ⟪ (a ◁ m) ⊔ γ ⟫ ⟶ [send a, b(⦃ m ⦄pk(b))] B a, (X b, m)
| OSS₂ : Axiom $ ⟪ γ ⟫ ⟶ [recv b(m)]B b, (X a, m)
\end{lstlisting}

The first axiom states that, from a protocol configuration in which the initiator knows the message it intends to transmit, after it actually sends the message to the responder, the initiating agent believes that the newly transmitted message becomes explicitly known to the other agent. The second axiom studies the epistemic relation from the perspective of the receiving agent.
We can now state and prove the following theorem: starting from the initial protocol state, after the  message transmission, the responder knows that the initiator explicitly knows the sent message; this proof constitutes part of the mutual authentication claim. 

\begin{lstlisting}
def OSS : ProtocolsProof _ $ ⟪ (i ◁ n) ⊔ γ₀ i r ⟫ ⟶ [(send i, r(⦃ n ⦄pk(r)))]([(recv r(n))](K r, (X i, n)))
\end{lstlisting}

\subsection{The Modal Fragment. S5 Reasoning}
Finally, we show that whenever the full expressive power of the logic is unneeded, one may use the same underlying formalism to reason about a less expressive fragment. To this end, we describe a \textit{shallow embedding} of plain modal logic onto our \textit{deep embedding} of many-sorted hybrid logic in Lean. We give a simple definition of the basic modal language:

\begin{lstlisting}
hybrid_def ModalBase :=
  sort WFF ::= "◇"(WFF) [poss]
\end{lstlisting}

This declares a many-sorted signature \lstinline{ModalBase}, containing a single (unary) operator \lstinline{◇}, the modal diamond. The operator \lstinline{□} is defined as syntax macro for the dual of \lstinline{◇}.

As is, this signature is that of \textit{regular hybrid logic}, mono-sorted and monadic. We want, however, to go further and enforce the restriction that \textit{no state symbols appear} in the modal fragment whatsoever, removing all hybridization from the language.
The way to do this is straightforward. A fragment of the language is \textit{a boolean predicate} on formulas (parametrized by the signature of the language):

\begin{lstlisting}
def Fragment (symbs : Symbols α) :=
  (ss : List symbs.signature.S) → FormL symbs ss → Bool
\end{lstlisting}

We define \lstinline{IsBase}, the basic modal fragment, as \lstinline{false} for all hybrid constructors:
\begin{lstlisting}
def IsBase : Fragment ModalBase
  | _, .prop _   => true
  | _, .appl _ φ => IsBase _ φ
  | _, .or φ ψ   => IsBase _ φ && IsBase _ ψ
  | _, .neg φ    => IsBase _ φ
  | _, _         => false
\end{lstlisting}

We define \lstinline{Modal} as the type of formulas in the signature \lstinline{ModalBase}, which additionally come with a proof of \lstinline{IsBase}. Next, a \textit{proof} is in a certain syntactic fragment if \textit{all its component subproofs} only reference formulas within that fragment. We express this by the following predicate, of which we show only the type, for brevity (the definition is by structural recursion on the proof):

\begin{lstlisting}
def Proof.inFragment (P : Fragment symbs) : Proof Λ s φ → Bool :=
\end{lstlisting}

So how can we define, say, S5? As before, we introduce the respective axiomatic extension (note that axiom K is not needed because it is already part of the primitive proof system):

\begin{lstlisting}
inductive S5Schema : Modal → Type where
  | AxT : S5Schema (□ φ ⟶ φ)
  | Ax5 : S5Schema (◇ φ ⟶ □◇φ)
def S5 : AxiomSet ModalBase :=
  λ _ => setOf ( λ form => ∃ (φ : Modal) (_ : S5Schema φ), φ.toForm ≍ form )
\end{lstlisting}

An S5 proof is a regular proof from the axioms \lstinline{S5}, which is \textit{strictly in the basic modal fragment}. Formally:

\begin{lstlisting}
def S5Pf (φ : Modal) := Proof.fragment IsBase S5 φ.1
\end{lstlisting}

Putting everything together, this embeds the syntax and proof system of S5 in many-sorted hybrid polyadic logic. To this extent, we display a sample S5 proof:
\begin{lstlisting}
def posNecActual : S5Pf (◇□ φ ⟶ φ) := by
  have l1 : S5Pf (◇(~φ) ⟶ □◇(~φ)) := ax_a5
  have l2 : S5Pf (~(□◇(~φ)) ⟶ ~(◇(~φ))) := modusPonens contraP l1
  have l3 : S5Pf ((◇□φ) ⟶ ~(□◇ ~φ)) := dni
  have l4 : S5Pf ((~(□◇(~φ)) ⟶ ~(◇(~φ))) ⟶ (◇□φ ⟶ ~(◇(~φ)))) := modusPonens impTrans l3
  have l5 : S5Pf (◇□ φ ⟶ □ φ) := modusPonens l4 l2
\end{lstlisting}
\vspace{-1.25em}
\begin{lstlisting}
  have l6 : S5Pf ((□ φ ⟶ φ) ⟶ (◇□ φ ⟶ φ)) := modusPonens impTrans l5
  have l7 : S5Pf (◇□ φ ⟶ φ) := modusPonens l6 ax_t
  exact l7
\end{lstlisting}

\section{Related Work and Conclusions}\label{related}

Formalizations of modal and hybrid logics in proof assistants have attracted sustained attention over the last decade. In the hybrid settings, \cite{hybridisabelle} provides a formalization of a Seligman-style tableau system for hybrid logic in Isabelle/HOL, including soundness and completeness results, while \cite{hybridcoq} develops a constructive formalization of hybrid logic in Rocq. More recently, \cite{itpagents24} presents a Lean formalization of the completeness proof for coalition logic with common knowledge, proving that canonical-model constructions can be mechanized in Lean. Our work differs from these developments since, first, we target a many-sorted polyadic hybrid modal logic, that significantly increases the complexity of both the syntax compared to mono-sorted settings; and since our emphasis lies on a uniform representation of signatures and axiomatic extensions that can be instantiated to concrete systems, rather than on a single fixed logic. In parallel, research on many-sorted logics has evolved along different lines, for example Matching $\mu$-logic \cite{matching} offers a highly expressive framework for reasoning about program configurations and structural properties; its mechanization in Rocq \cite{matchingcoq} and the subsequent many-sorted Lean development \cite{matchinglean} (for Applicative Matching Logic) show that sort-parametric reasoning can be effectively integrated into proof assistants. We also refer to \cite{loom} for a Lean-based framework used for  verifying executable semantics.



The logical system mechanized in this work provides a rigorous foundation for future extensions. As future work we intend to prioritize: (1) the mechanization of the Completeness Theorem, as stated in \cite{tableaux}; (2) further support for Hoare-style reasoning, including loops and proof automation via VCG; and (3) devising tactics and constructions that facilitate the interactions with various users that might not be familiar with the details of our implementation.

\bibliography{lipics-v2021-msphl-article}

\end{document}